\documentclass[aps,twocolumn,showpacs,showkeys,superscriptaddress]{revtex4}
\usepackage{amsmath}
\usepackage{amssymb}
\usepackage{graphicx}
\newcommand{\pd}{\partial}
\newcommand{\tr}{s}

\newcommand{\nn}{\nonumber}
\begin{document} 
\title{Multiply Phased Traveling BPS Vortex} 
\author{Kyoungtae Kimm}
\email{helloktk@snu.ac.kr}
\affiliation{Faculty of Liberal Education, 
Seoul National University, Seoul 151-747, Korea}
\author{J. H. Yoon}
\email{yoonjh@konkuk.ac.kr}
\affiliation{Department of Physics, Konkuk University, Seoul 143-701, Korea}
\author{Y. M. Cho}
\email{ymcho7@konkuk.ac.kr}
\affiliation{Administration Building 310-4, Konkuk University, Seoul 143-701, Korea}
\affiliation{School of Physics and Astronomy, Seoul National University,
Seoul 151-742, Korea}
\begin{abstract}
We present the multiply phased current carrying vortex solutions 
in the U(1) gauge theory coupled to an $(N+1)$-component SU(N+1) 
scalar multiplet in the Bogomolny limit. Our vortex solutions 
correspond to the static vortex dressed with traveling waves along 
the axis of symmetry. What is notable in our vortex solutions is 
that the frequencies of traveling waves in each component of the scalar 
field can have different values. The energy of the static vortex is 
proportional to the topological charge of $CP^N$ model in the BPS 
limit, and the multiple phase of the vortex supplies additional 
energy contribution which is proportional to the Noether charge 
associated to the remaining symmetry.
\end{abstract}

\maketitle

\section{Introduction}

Topological string-like vortex solutions have been the subject 
of intense studies in field theory and cosmology since 
the discovery of the Nielsen-Olesen (NO) string in the Abelian
Higgs model \cite{ano}. Their existence is closely connected 
to the spontaneous breaking of the U(1) gauge symmetry by Higgs 
mechanism. The string-like solutions have been shown to exist 
when the theory is generalized to have a global SU(2) symmetry 
with a scalar doublet \cite{vacha91,hind91}. This is interesting 
because this model represents the bosonic sector of the electroweak 
theory in the limit that the Weinberg angle, $\sin^2 \theta_\text{W}$, 
becomes one.

One of the most interesting properties of this solution is 
that its stability is guaranteed without being supported by 
the standard topological argument related to the homotopy 
of the vacuum manifold. In the Bogomolny-Prasad-Sommerfield
(BPS) limit where the mass ratio of gauge field and the Higgs 
become equal, there is a continuous family of solutions which 
are degenerate in energy \cite{bogo}. The solution space 
consists of either the embedded NO vortices or the baby 
skyrmion solutions.
 
In recent years, new type of solutions were found in this 
model. These twisted ``semilocal" vortex were constructed 
allowing the two scalar components to have different phase 
factors which depend on time as well as on the coordinate 
corresponding to their axis of symmetry \cite{forga06}. 
A notable fact about the twisted vortex is that the energy  
away from the BPS limit can be lower than the energy of
the corresponding embedded NO vortex.

Clearly this type of vortex solutions is very interesting 
from the mathematical point of view. Moreover, from 
the physical point of view they also become interesting 
because they have potentially important applications in 
various areas of physics. Obviously they could play 
important roles in condensed matter physics, in particular 
in multi-gap superconductors and multi-component 
Bose-Einstein condensates \cite{pra05,epjb08}. This is 
because they are a natural generalization of the Abrikosov 
vortex in Ginzburg-Landau model of superconductor.  

Moreover, they may have a natural application to the Skyrme 
theory, because the Skyrme theory also has the global SU(2) 
and local U(1) symmetry \cite{ijmpa08}. And they could play 
important roles in high energy physics because they could 
be embedded in the standard Weinberg-Salam model. Finally, 
they could describe cosmic strings and become important 
in cosmology. So they have interesting applications in 
almost all areas of physics.

The aim of this paper is to show that the multiply phased 
vortex solutions exist when we generalize the global SU(2) 
to SU(N+1). Especially we find the multiply phased vortex 
solutions in the BPS limit. The solutions can be obtained 
by dressing the traveling wave which moves along the axis 
of symmetry. What is remarkable in our vortex solutions is 
that the frequencies of traveling waves in each component 
of the SU(N+1) scalar multiplet can be different, so that
they are multiply phased. 

The energy of the static BPS vortex is proportional to 
the topological charge of the $CP^N$ model, and the multiple 
phase (or the ``twisting") of vortex supplies additional 
energy contribution which is proportional to the Noether 
charge associated to the remaining symmetry. We have analyzed 
this additional energy contributions in terms of known static 
BPS solutions and electromagnetic charge density in details.

The paper is organized as follows. In Section II we investigate  
the static BPS semilocal vortex solutions when the scalar field 
has $N+1$ components. We express all relevant equations in 
a gauge invariant way. In Section III we construct the multiply 
phased vortex solutions which have traveling waves along their 
axis of symmetry. We show the energy of multiply phased BPS 
vortex can be written in terms of topological charge of $CP^N$ 
model and Noether charge. In the final section we discuss 
the physical implications of our results. In this paper we 
use the metric where $g_{\mu\nu} = \text{diag}(+1,-1,-1,-1)$.

\section{Static BPS Vortex}

The so-called ``semi-local" string appears naturally in models 
involving multiply charged complex scalar fields coupled to 
electromagnetism which has a spontaneous symmetry breaking. 
They were introduced as a minimal extension of the Abelian 
Higgs model with the familiar topologically stable 
Nielsen-Olesen vortex. 

In this extended model, the Abelian gauge field couples to 
a complex Higgs doublet so that in addition to local U(1) 
symmetry, there is an extra SU(2) global symmetry, which 
is spontaneously broken to a global U(1). According to 
the popular wisdom the trivial first homotopy of vacuum 
manifold (i.e., $\pi_1(S^3)=0$) rejects the existence of 
topological vortex, but when the scalar field at the spatial 
infinity lies on a gauge orbit or a circle lying in the vacuum 
manifold, the gradient energy of scalar field could have 
a finite value and thus the theory could support finite
energy solitons. This leads to U(1) vortex solutions even 
though the vacuum manifold is simply connected. They 
have important applications in cosmology, multi-component 
superconductor in condensed matter physics \cite{sigrist}, 
and in two-component plasma physics \cite{lubcke}. 

Recently, it was shown that the semilocal vortex can carry 
persistent longitudinal currents associated with the global 
symmetry subgroup \cite{forga06}. Our primary interest 
in the model is these properties of BPS semilocal vortex 
solutions. We found that it is possible to have a vortex 
with traveling wave whose frequencies are all different 
when the scalar field has more than two components.

Consider the extended Abelian Higgs model in which an SU(N+1) 
multiplet scalar field $\phi = (\phi_1, \phi_2,...,\phi_{N+1})$ 
is coupled to the Abelian gauge field $A_\mu$,
\begin{align}
\label{lag1}
{\cal L} =  \frac{1}{2}|D_\mu \phi|^2 
+ \frac{\lambda}{8} (|\phi|^2-v^2)^2
-\frac{1}{4} (F_{\mu\nu})^2,
\end{align}
where $D_{\mu }\phi = \partial_\mu \phi +i e A_\mu \phi$ 
and $F_{\mu\nu}=\pd_\mu A_\nu - \pd_\nu A_\mu$ is 
the Abelian field strength.

The Lagrangian possesses the local U(1) gauge symmetry, 
being invariant under the transformations, 
$\delta \phi = i\alpha\phi, \delta A_\mu = -(1/e) \pd_\mu  \alpha$, 
where $\alpha(x)$ is an an infinitesimal parameter. 
Moreover, it has the global SU(N+1) symmetry 
$\delta \phi = i \alpha_a T_a \phi$, where $T^a ~(a=1,...,N+1)$ 
are the generators of SU(N+1) in the fundamental representation, 
and $\alpha_a$ are constant infinitesimal parameters.

When the scalar get a vacuum expectation values $v$ 
the gauge symmetry is spontaneously broken.
In the symmetry broken phase, the particle spectrum 
consists of a massive vector field with mass $m_A= ev$, 
and of scalar Higgs with mass $m_H = \sqrt{\lambda}v$. 
The remainings of the particle spectrum are $2N$-massless 
Goldstone bosons. For convenience we will rescale the fields 
so that all variables become dimensionless,
\begin{gather}
\phi \rightarrow v\phi,
~~~A_\mu\rightarrow vA_\mu,  \nn\\
x_\mu \rightarrow x_\mu/(ev),  
\end{gather}
in the following.

Introducing $CP^{N } $ field $\xi$,
\begin{align}
\phi = \rho \xi,\quad (\xi^\dagger \xi = 1),
\end{align}
the Lagrangian (\ref{lag1}) can be written as
\begin{gather}
\label{lag2}
{\cal L} = \frac{1}{2}(\pd_\mu \rho)^2 
+ \frac{1}{2}\rho^2 |D_\mu \xi|^2 -\frac{\beta}{8} (\rho^2 -1)^2  \nn\\
-\frac{1}{4} (F_{\mu\nu})^2,
\end{gather}
where now the covariant derivative becomes 
$D_\mu \xi = (\pd_\mu + i A_\mu )\xi$ and 
$\beta=\lambda /e^2$. In terms of $\rho$ and $\xi$ 
the conserved current associated with the local $U(1)$ 
gauge symmetry  can be written as 
\begin{gather}
\label{emcurrent}
J_\mu = \frac{1}{2i } \rho^2 ( \xi^\dagger D_\mu \xi 
- (D_\mu \xi )^\dagger \xi ) \nn\\
= \rho^2 (a_\mu +   A_\mu ),
\end{gather}
where $a_\mu =-i \xi^\dagger \pd_\mu \xi$ is the auxiliary 
U(1) gauge field of $CP^{N}$ model. With this $J_\mu$ which is 
gauge invariant we can express our equations in a gauge 
independent way.

Since 
\begin{gather}
|D_\mu \xi |^2 = |\nabla_\mu \xi |^2 
+\frac{ J_\mu^2 }{  \rho^4},   \nn\\ 
\nabla_\mu \xi = (\pd_\mu -i a_\mu )\xi 
= (\pd_\mu - \xi^\dagger \pd_\mu \xi) \xi,
\end{gather}
we can rewrite the Lagrangian in terms of gauge invariant 
quantities \cite{prd96},
\begin{gather}
{\cal L } = \frac{1}{2} (\pd_\mu \rho)^2 
+\frac{1}{2} \rho^2 |\nabla_\mu \xi|^2
+\frac{ J_\mu^2}{2 \rho^2 } -\frac{\beta}{8} (\rho^2 - 1)^2  \nn\\ 
-\frac{1}{4} (F_{\mu \nu})^2.
\end{gather}
The equations of motion  are then given as
\begin{gather}
\pd^\mu F_{\mu \nu} = -J_\nu,
\label{Aeq}   \\
\pd_\mu^2 \rho - |\nabla_\mu \xi|^2 \rho 
-\frac{ J_\mu^2 }{ \rho^3}
+\frac{\beta}{2} (\rho^2 -1)\rho = 0,
\label{rhoeq2}   \\
\nabla_\mu^2 \xi + 2\Big( \frac{\pd_\mu\rho}{\rho} 
+ i \frac{J_\mu }{ \rho^2}\Big)\nabla_\mu\xi 
= (\xi^\dagger \nabla_\mu^2 \xi) \xi.
\label{xieq2}
\end{gather}
The energy momentum tensor can be obtained varying 
the Lagrangian with respect to the metric tensor 
$g_{\mu\nu}$,
\begin{gather}
T_{\mu\nu} =  (\pd_\mu\rho)(\pd_\nu\rho) 
+ \frac{1}{2} (\nabla_\mu\xi)^\dagger \nabla_\nu \xi
+ \frac{1}{2}(\nabla_\nu\xi)^\dagger  \nabla_\mu\xi \nn\\
+\frac{J_\mu J_\nu}{\rho^2}
- F_{\mu\kappa}F_{\nu\kappa} -g_{\mu\nu}{\cal L}.
\end{gather}

It is well-known that this model with scalar doublet can have 
finite energy (per unit length) vortex solutions called 
the semilocal vortices. Some of the solutions in this model 
are just those obtained from the Nielsen-Olesen solutions 
identifying, say, the upper component of $\phi$ with 
the scalar field in the Nielsen-Olesen model. But there are 
different kinds of solutions. Especially, at the critical coupling 
there exist a solution which appears to describe a hybrid 
of a NO vortex and $CP^1$ lump in addition to the embedded 
NO solution \cite{hind91}.

Now, we review  the static straight vortex solutions whose 
axis is perpendicular to $(x_1,x_2)$ plane. The energy 
(per unit length) of static vortex configuration reads 
\begin{gather}
E = \int d^2x \Big[ \frac{1}{2}(\pd_i \rho)^2 
+\frac{1}{2} \rho^2 |\nabla_i \xi |^2   
+ \frac{\beta}{8}(\rho^2- 1)^2 \nn\\
+ \frac{ J_i^2}{2  \rho^2}+\frac{1}{2}B^2\Big],
\end{gather}
with $i=1,2$ and $B=F_{12}=\epsilon_{ij}\pd_i A_j$ is 
the component of magnetic field perpendicular to $(x_1,x_2)$ 
plane. If one now chooses the coupling constants to satisfy
\begin{align}
\beta =\frac{\lambda}{e^2}= 1,
\end{align}
the we can find a bound for the energy. At this critical value 
the two mass scales in the theory, scalar mass 
$m_H =\sqrt{\lambda} v$ and gauge field mass $m_A = ev$, 
are equal. This defines the Bogomolny limit in the Abelian 
Higgs model \cite{bogo}.

In order to show the Bogomolny limit in the current model, 
we will start by rearranging the term in the energy functional 
as follows
\begin{gather}
\frac{1}{2}(\pd_i\rho)^2 +\frac{ J_i^2 }{2 \rho^2}
=\frac{1}{2}\Big(\frac{J_i}{ \rho}  \pm  \epsilon_{ik} \pd_k \rho\Big)^2
\mp \epsilon_{ik} J_i \pd_k\ln \rho,
\nonumber \\
\frac{1}{2} |\nabla_i \xi|^2 =
\frac{1}{4} |\nabla_i \xi  \pm 
i \epsilon_{ik} \nabla_k \xi |^2  \pm  \frac{1}{2} f_{12},
\nonumber \\
\frac{B^2}{2} + \frac{1}{8} (\rho^2 - 1)^2
=\frac{1}{2} \Big[B  \mp  \frac{1}{2}(\rho^2-1) \Big]^2 
\mp \frac{1}{2}  B  \nn\\
\pm \frac{1}{2}\pd_i (\epsilon_{ik} J_k) 
\mp  \epsilon_{ik} J_i \pd_k\ln \rho
\mp \frac{\rho^2}{2} f_{12},
\nonumber
\end{gather}
where $f_{\mu\nu} = \pd_\mu a_\nu - \pd_\nu a_\mu$ is 
the field strength of the auxiliary $CP^N$ gauge field $a_\mu$.

After making use of this relations the energy of vortex can
be written as
\begin{gather}
E = \int d^2 x \Big\{ \frac{1}{2} \Big( \frac{J_i}{  \rho} 
\pm \epsilon_{ik} \pd_k \rho \Big)^2
+\frac{1}{4}\rho^2 |\nabla_i \xi 
\pm  i \epsilon_{ik} \nabla_k \xi|^2 \nn\\
+\frac{1}{2}\Big( B \mp \frac{1}{2}(\rho^2- 1 )  \Big)^2
\mp \frac{1}{2}  B \Big\}.
\label{eng2}
\end{gather}
From the last expression, we secure the Bogomolny bound
\begin{align}
E \ge \frac{1}{2}|\Phi|, \qquad ( \Phi=\int d^2x B),
\end{align}
since the we can choose upper sign or lower sign in (\ref{eng2}), 
depending on the sign of the flux $\Phi$.
For a given value of $\Phi$, this bound is saturated if and only if 
the fields satisfy the self-duality equations
\begin{gather}
\label{selfdualCPN}
\nabla_i \xi  \pm 
 i \epsilon_{ik} \nabla_k \xi =0 ,\\
\label{currenteq}
J_i  \pm 
 \frac{1}{2} \epsilon_{ik} \pd_k \rho^2   = 0 , \\
\label{mageq}
B  \mp \frac{1}{2} (\rho^2  -1)  = 0.
 \end{gather}
Any solution to the first order self-duality equations automatically 
satisfy the original static field equations. In what follows we shall 
focus on vortex solutions (the upper sign)  without loss of 
the generality.

Let us now discuss about the nature of these self-dual soliton 
solutions. The last two equations (\ref{currenteq}) and (\ref{mageq}) 
can be combined to give a single second order differential equation
\begin{align}
\pd_i^2 \ln {\rho }
-\frac{1}{2}(\rho^2 -1)
-  f_{12} = 0.
\label{heq2}
\end{align}
Note that the topological charge of $CP^N$ model is defined by
\begin{align}
T = \frac{1}{2\pi i} \int d^2x 
\epsilon_{ij} ( \nabla_i \xi)^\dagger\nabla_j \xi = \frac{1}{2\pi }
\int d^2x  f_{12}.
\end{align}
For an embedded NO vortex solution the scalar field $\rho$ should 
vanish at the location of vortex in order to have well defined phase. 
In this case we have
\begin{gather}
f_{12} = 2\pi \sum_{r=1}^{n} \delta (\vec{x}-\vec{x}_r),  \nn
\end{gather} 
where $\{\vec{x}_1, \vec{x}_2,..., \vec{x}_n  \}$ are the locations of 
vortex core on the plane.
 
Thus the magnetic flux of NO vortex measures the vorticity
\begin{align}
\int d^2 x  B = 2\pi n.  \nn
\end{align}
When $f_{12}=0$, the solution of Eq.~(\ref{heq2}) describes 
a embedded self-dual NO vortex configuration. In this case 
the scalar field $\rho$ has to vanish at the origin in order to 
have a well defined phase of vortex solution. Near the vortex 
cores we have 
\begin{gather}
\pd_i^2 \log \rho = 2\pi n \delta^2(\vec{r}),   \nn
\end{gather}
for $n$ superimposed vortex. Furthermore, with (\ref{emcurrent}) 
and  (\ref{currenteq}) we can show that
\begin{align}
\int d^2 x e B 
& = \int d^2x ~\pd_i^2 \log \rho - \int d^2x f_{12}  
\nonumber \\
& =2\pi r \frac{d\ln \rho}{dr}\Big|_0^\infty - 2\pi T,
\end{align}
Since  $\rho(r) =\rho_0  r^n +...$ for the embedded NO vortex,
the magnetic flux has quantized value $|e\int d^2x B |=2\pi n$.
When the scalar field $\rho$ has no zeros, then the magnetic flux 
is solely expressed in terms of  the topological charge of $CP^N$ 
model, 
\begin{align}
 \int d^2x B =2\pi T.  
\end{align}
There are two kind of degenerate solutions depending on 
the value of scalar field $\rho$ at the origin, embedded NO vortex
and dubbed baby skyrmions \cite{vacha91, hind91}. 
From now on we will concentrate on the self-dual  baby skyrmion.

\begin{figure}
\includegraphics[height=4cm]{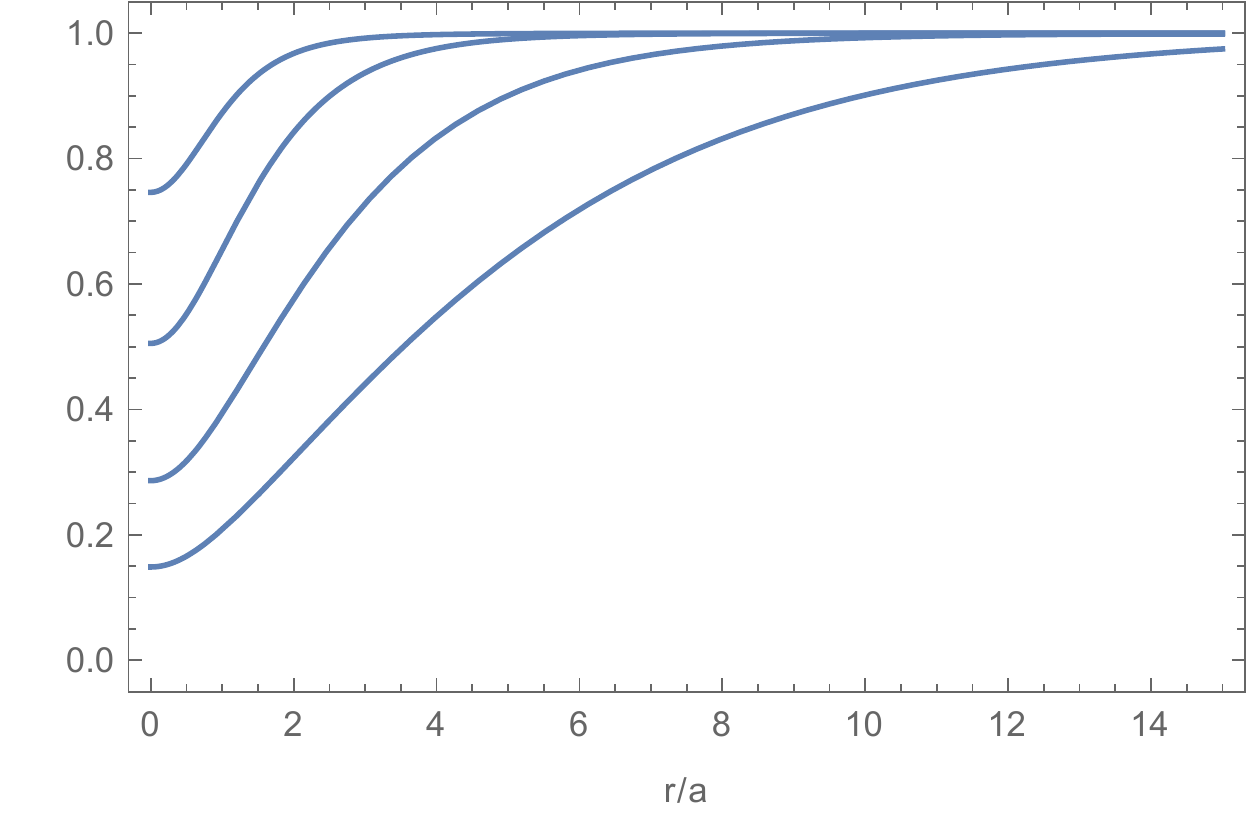}
\label{rho1}
\caption{\label{rho1} Profiles of $\rho$ for $k=1$ with $a=0.25, 0.5, 1,2$. 
Notice that $\rho(0)$ is monotonically increasing as $a$ becomes 
larger.}
\end{figure}

The most general solution for Eq. (\ref{selfdualCPN})  can be 
given in terms of $N$ (anti-)holomorphic functions 
$u_a~(a=1,2,...,N)$,
\begin{align}
\xi =\frac{1}{\sqrt{1+ |u|^2} } 
\left(\begin{array}{c}
u \\
1
\end{array}\right).
\end{align}
where $u=(u_1,u_2,...,u_N)^T$ is $N$ column vector.
The topological charge density can be written as 
\begin{gather}
f_{12} 
=  \frac{1}{2}\pd_i^2 \ln (1+ |u|^2) 
=\frac{2 (u')^\dagger \Delta ^2 u'} {(1+ |u|^2)^2} \nn\\
=\frac{2 |\Delta  u'|^2} {(1+ |w|^2)^2} ,
\end{gather}
where $u'$ means the derivative of $u$ with respect to 
its argument and $\Delta^2 $ is a $N\times N$ matrix 
defined by
\begin{align}
\Delta^2 = (1+ |u|^2) I_N  - u \otimes u^\dagger,
\end{align}
where $I_N$ is a $N\times N$ identity matrix.

Let us consider the simplest $CP^1$ rotationally symmetic 
lump solution, $u=u_1 = (a/z)^k$ where $z=x_1+i x_2$ 
and $a$ is the width parameter  of soliton. Since $CP^1$ 
model is scale invariant, the width parameter $a$ cannot be fixed.
The topological charge of this lump solution is given as $T=k$, 
and $\Delta^2 = 1$. With a  radial coordinate $\tr=  r/ a$ rescaled 
with respect to the width parameter $a$ of $CP^1$ lump, 
the condensate equation (\ref{heq2}) becomes
\begin{align}
\frac{1}{ \tr}\frac{d}{d\tr }\Big(\tr \frac{d}{d\tr} \ln  \rho \Big)
-\frac{a^2}{2}   (\rho^2-1)
= \frac{2k^2 \tr^{2k-2}}{ (1 + \tr^{2k})^2}
\end{align}
Although solution in the BPS limit is characterized with parameter 
$a$ the energy is independent of the width of solution. This implies 
the existence of  a zero model associated with the width of semilocal
vortex.

For small values of $s$, 
\begin{align}
 \rho(\tr) = b\Big[ 1 + \Big( \frac{\delta_{k1} }{2} 
 - \frac{a^2}{8}+\frac{a^2b^2}{8}\Big)\tr^2   +...\Big]. 
\end{align}
For a given $CP^1$ lump size $a$,  we can find the solution $\rho$ 
which goes to the vacuum as $\tr\rightarrow\infty$ with appropriate 
choice of $\rho(0)=b$ as a shooting parameter. 
The asymptotic behavior of condensate $\rho$ is 
\begin{align}
\rho(\tr) = \Big(1-\frac{2}{\tr^{2k}} +...
+C \frac{e^{-a\tr}}{\sqrt{\tr}}+...\Big),
\end{align}
Fig. \ref{rho1} and Fig. \ref{rho2} show the profiles of 
$\rho$ for $k=1,2$ for various width parameter values.

Since the magnetic field is related with the $\rho$ by 
Eq. (\ref{mageq}), we can see that magnetic field will show 
power-law decay asymptotically. This feature is in sharp 
contrast with the embedded NO vortex where the magnetic 
field is confined within the size determined by the mass of 
gauge field.

\begin{figure}
\includegraphics[height=4cm]{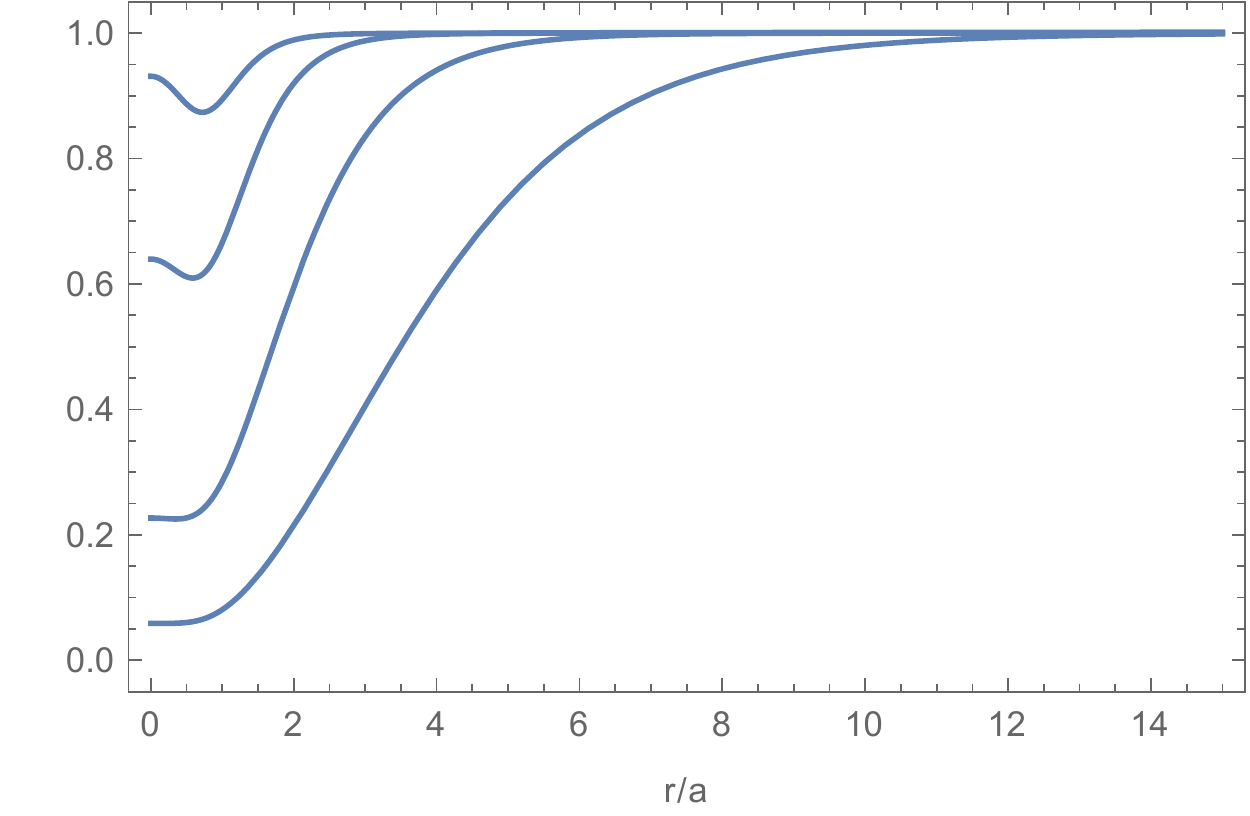}
\label{rho2}
\caption{\label{rho2} Profiles of $\rho$ for $k=2$, with 
$a=0.5,1, 2, 4$. Here again $\rho$ increases monotonically.}
\end{figure}

\section{Multiply Phased BPS Vortex}

Now, let us investigate  semilocal vortex solutions with 
space-time dependent phases. Requiring vortex solution 
to have translational symmetry in time and along the vortex 
axis, $x_3$, the scalar fields can get a phase which is linearly 
dependent on $t$ and $x_3$. In particular, we are interested 
in vortex solutions with traveling wave type phases along 
its axis $x_3$, 
\begin{gather}
\phi = \rho(r) \xi,
\nonumber \\
\xi =\frac{1}{\sqrt{1+ |u|^2} } 
\left(\begin{array}{c}
 u_1(z)e^{i \omega_{1}( t - x_3)} \\
u_2(z)e^{i \omega_{2}( t - x_3)} \\
\vdots \\
u_N(z)e^{i \omega_{N}( t - x_3)}
\\
1
\end{array}\right), 
\nonumber \\
 A_\mu  = A_\mu (x_1, x_2),
\label{twistsol1}
\end{gather}
where 
the gauge field is assumed to have no $t,x_3$-dependency.
Note that the frequencies of traveling wave in each component 
of scalar field are all distinct values. Thus there are relative 
phases (twist of phase) between any two components of scalar 
field. This feature cannot be realized in the semilocal model
with scalar doublet. A similar construction of $CP^N$ lump 
solutions dressed with traveling waves in the ungauged 
$CP^N$ model has been considered in \cite{ferreira2011}.

This $t$ and $x_3$ dependency in twisted vortex ansatz gives 
nontrivial  $a_0$ and $a_3$ components of the auxiliary field:
\begin{gather}
\label{a0a3}
a_0 = -a_3 = \frac{u^\dagger \Omega u}{1+|u|^2},
\end{gather}
where
\begin{align}
\Omega =  \text{diag}(\omega_1, \omega_2, ..., \omega_N).
\end{align}
Again this permits a twisted vortex to have 
  nonzero charge density $J_0$ and a longitudinal current 
density $J_3$, which is obvious from (\ref{emcurrent}).

With the relations
\begin{gather}
(\nabla_0 +\nabla_3 )\xi = 0,   \nn\\
(\nabla_0^2 - \nabla_3^2 )\xi=0,
\end{gather}
 one can show that 
the twisted vortex ansatz becomes a solution of equations 
of motion (\ref{Aeq}), (\ref{rhoeq2}), and (\ref{xieq2})  when
the following condition is met,
\begin{align}
J_0 = -J_3.
\end{align}
With this relation the self-duality equations (\ref{selfdualCPN}), 
(\ref{currenteq}), and (\ref{mageq}) defined in $(x_1, x_2)$ plane 
are completely decoupled from the those in $(t, x_3)$ directions.
If the wave were traveling down along the $x_3$-axis, then we 
should have $J_0=+J_3$. Both the cases satisfy the relation 
$J_0^2 = J_3^2$. This implies that there is a longitudinal 
electric current $J_3 $ carried by traveling waves and the region 
with nonvanishing current is electrically charged.

From (\ref{Aeq}) we can see that  the charge density $J_0$ 
is governed by the following equation,
\begin{align}
\label{j0eq}
\pd_i^2 \Big( \frac{J_0}{\rho^2 } \Big)  - \pd_i f_{i0} =   J_0.
\end{align}
The second term can be expressed in terms of $u$ explicitly:
\begin{gather}
\partial_i f_{i0} = \frac{4(u')^\dagger \Delta^2 \Omega u '}{(1+|u|^2 )^2}
-\frac{4(u')^\dagger \Delta \Omega u}{(1+ |u|^2)^3}(u^\dagger  u') 
\nonumber \\
-\frac{4(u')^\dagger \Delta ^2 u}{(1+ |u|^2 )^3} (u^\dagger \Omega u).
\end{gather}
Thus with the solution of (\ref{heq2})
we can solve Eq.~(\ref{j0eq}) in principle. 
 
If all traveling waves have the same frequency, it will give a much 
simpler vortex solution. For generality, we will discuss solutions  
in which each traveling wave in the $a$-th $CP^N$ component 
has a distinct frequency.

Note, the left hand side of (\ref{j0eq}) can be written 
\begin{gather}
\pd_i^2 (\dfrac{J_0}{\rho^2}- a_0) = J_0.
\end{gather}
Integrating both sides of this  
over the transverse plane gives vanishing  net charge per unit 
vortex length,
\begin{align}
q_\text{tot} = \int d^2x J_0 = 0
\end{align}
This also implies the vanishment of net longitudinal current because of 
the relation  $J_0^2= J_3^2$.

The ansatz (\ref{twistsol1}) breaks the original global SU(N+1) 
symmetry to N U(1). Under this residual global symmetry 
the fields $\xi$ transform as
\begin{gather}
\label{ressym}
\delta \xi_a  = {i \alpha_a} \xi_a,~~a=1,2,...,N,
~(\text{no sum over $a$}),
\nonumber \\
\delta \xi_{N+1}  = 0.
\end{gather}
The Noether currents assoicated with these symmetries are given by 
\begin{align}
{\cal J}^{(a)}_{\mu} =   \rho^2 \big(  \xi_a^\dagger (D_\mu \xi)_a^{}
- (D_\mu \xi)_a^\dagger \xi_a^{} \big),
\end{align}
where no summation is implied over $a=1,2,..., N$.
Since $\pd_0 \xi_a = i \omega_a \xi_a$ and $\pd_0 \xi_{N+1}=0$,
we get the following relation
\begin{align}
 \sum_{a=1}^N \omega_a {\cal Q}^{(a)}  
=\int d^2x  \Big( \rho^2 |\nabla_0\xi |^2 + a_0 J_0 \Big),
\end{align}
where ${\cal Q}^{(a)}$ is the Noether charge per unit length of the vortex
\begin{align}
{\cal Q}^{(a)} = \int d^2x {\cal J}_0^{(a)}.
\end{align}
In the following we will show that the twisted vortex carry  
these global Noether charges.

The tension of twisted vortex (energy per unit length) is given by
\begin{gather}
{E} =\pi  T + E_\text{twist}, \nn\\
E_\text{twist} = \int d^2x\bigg(\frac{1}{2}\rho^2|\nabla_0\xi|^2 
+ \frac{1}{2}\rho^2|\nabla_3 \xi|^2  \nn\\
+\frac{J_0^2}{2 \rho^2} +\frac{J_3^2}{2 \rho^2}
+\frac{1}{2} E_i^2+ \frac{1}{2} B_i^2 \bigg),
\end{gather}
where  $E_i = F_{i0}$ and $B_i = \epsilon_{ij}F_{j3},~(i,j=x_1,x_2)$
are the transverse components of electric and magnetic field 
respectively.  From (\ref{emcurrent}) we have 
\begin{gather}
E_i = \pd_i \Big(\frac{J_0}{\rho^2} - a_0\Big), 
= -\epsilon_{ij} E_i,
\label{efield1}
\end{gather}
and with (\ref{j0eq}) the energy of transverse components 
of electric field  can be expressed as
\begin{align}
\label{tedensity}
\int d^2x  E_i^2 =  \int d^2x \Big(  -\frac{J_0^2}{\rho^2} + a_0 J_0 \Big).
\end{align}
Since $\pd_0 \xi  = - \pd_3\xi$ and $J_0 = -J_3$ the transverse 
components of magnetic field are given as
\begin{align}
B_i = -\epsilon_{ij} E_j.
\end{align}
Thus the energy contribution from twisting of vortex  can be 
expressed with the Noether charges,
\begin{gather}
E_\text{twist} = \int d^2x\Big( \rho^2 |\nabla_0 \xi|^2 
+ \frac{J_0^2}{\rho^2} + E_i^2 \Big)   \nn\\
 =  \sum_{a=1}^{N} \omega_a {\cal Q}^{(a)} .
\end{gather}
For a embedded ANO vortex or skyrmion solutions correspond 
to untwisted ($\omega_k = 0$) static solutions.

This shows that the energy of a twisted BPS vortex of 
the type (\ref{twistsol1}) depends on  the global Noether 
charges in addition to  the topological charge,
\begin{align}
E  =  \pi T + \sum_{a=1}^N \omega_a {\cal Q}^{(a)} .
\end{align}
This is unlike the case of untwisted semilocal vortex 
where the Bogomolny bound is given in terms of 
the topological charge only.

Let us now discuss the topological properties of the $E_\text{twist}$.
$E_\text{twist}$ can be into three parts as have done in Ref. \cite{chernodub}. 
Each parts can be expressed with  scalar field $\rho$ determined from 
(\ref{heq2}), $CP^{N}$ lump solution $w(z)$, and charge density $J_0$ 
determined from (\ref{j0eq}).

With (\ref{efield1}) and (\ref{j0eq}) we have the following relation
\begin{align}
& \int d^2 x\Big( E_i^2 +\frac{J_0^2}{\rho^2} \Big) = 
\int d^2x \Big(\frac{J_0}{\rho^2} \pd_i f_{i0}  + f_{i0}^2 \Big).
\end{align}
We can obtain the similar relations for $J_3$, $f_{i3}$, and $B_i$.
Using these we have the following expression for the energy 
contribution from twisting. Thus the energy contribution from 
twisting consists of three separate contributions. 
\begin{align}
E_\text{twist} &=
\int d^2x \Big(  \rho^2|\nabla_0\xi|^2 
+ f_{i0}^2 
+ \frac{J_0}{ \rho^2}\pd_i f_{i0} \Big).
\end{align}
Firstly, note that 
\begin{align}
\epsilon_1 =\int d^2x \rho^2 |\nabla_0 \xi|^2
=  \int d^2x\frac{ |\Delta \Omega { u}|^2}{(1+|u|^2)^2} \rho^2.
\end{align}
where 
\begin{align}
\Omega =  \text{diag}(\omega_1, \omega_2, ..., \omega_N).
\end{align}
If the frequencies of all traveling waves are equal, 
$\Omega =\omega_0 I_N$, then we have 
$|\Delta \Omega{u}|^2 = \omega_0^2 |u|^2$ so that 
the integral is proportional to $\omega_0^2$. 
However, for a configuration with unit topological charge 
this term diverges logarithmically since as $r\rightarrow \infty$, 
$\rho \rightarrow \text{const}$, and $|u|\rightarrow 1/r$
as have noted in Ref. \cite{abraham}.

The second contribution is given by
\begin{align}
\epsilon_2=\int d^2x   f_{i0}^2 
=  \int d^2x  \frac{ |{u}^\dagger\Omega \Delta^2 u' |^2 }{(1+ |u|^2)^4 },
\end{align}
which can be evaluated if  a explicit form of $u$ were given. 
This integral is also proportional to $\omega_0^2$
if all frequencies of waves are equal to $\omega_0$. 
 
Remaining contribution to energy can be written as
\begin{gather}
\epsilon_3=  \int d^2x \frac{J_0}{\rho^2} \pd_i f_{i0}   \nn\\
 = \frac{4}{i} \int d^2x \bigg\{\frac{(u')^\dagger \Delta^2 
\Omega{u}'}{(1+ |u|^2)^2} -\frac{ ((u')^\dagger \Delta^2 
\Omega{u})(u^\dagger u')}{(1+ |u|^2)^3} \nn\\
-\frac{ |\Delta u'|^2 (u^\dagger \Omega{u})}{(1+|u|^2)^3}
\bigg\}\frac{J_0 }{\rho^2},
\end{gather}
where we used the relation (\ref{emcurrent}).  Again if all 
traveling waves has the same frequency $\omega_0$, 
then one can notice from Eq.~(\ref{j0eq}) that $\pd_i f_{i0}$ is 
proportional to $\omega_0$. This implies  that $J_0/\rho^2$ 
is proportional to $\omega_0$ and the last integral is proportional 
to $\omega_0^2$. Thus twisting the BPS semilocal vortex  
gives additional energy contributions which depend on 
the frequencies $\Omega$ as well as topological charge $T$:
\begin{align}
E= \pi   T + \epsilon_1(T, \Omega)+ \epsilon_2(T, \Omega)
+ \epsilon_3(T, \Omega).
\end{align}

Since the ansatz (\ref{twistsol1}) moves along the gauge orbit 
of the unbroken global $U(1)^N$ as one moves along 
the $x_3$ axis, $\pd_3 \xi_a= -i \omega_a \xi_a,~(a=1,2,...,N)$, 
we expect the translational symmetry along $x_3$ axis.
Consequently a twisted vortex should carry the conserved 
longitudinal momentum
\begin{gather}
P = \int d^2x T^{3}_0 \nn\\
=\int d^2x \Big( F_{0i}F_{3i}   
- \frac{\rho^2}{2}(\nabla_0\xi)^\dagger (\nabla_3\xi)   \nn\\
-\frac{\rho^2}{2} (\nabla_3\xi)^\dagger (\nabla_0 \xi) 
-\frac{J_0J_3}{\rho}\Big).
\end{gather} 
Indeed, it can be written in terms of the Noether charges as 
the corresponding conserved longitudinal momentum 
(per unit length) $P=T_0^3$ carried by 
a twisted vortex  also can be  expressed  in terms of Noether charges,
\begin{gather}
P=\int d^2x \Big( F_{0i}F_{3i} 
+ \frac{\rho^2}{2}(\nabla_0\xi)^\dagger (\nabla_3\xi)  \nn\\
+\frac{\rho^2}{2} (\nabla_3\xi)^\dagger (\nabla_0 \xi)
+\frac{J_0J_3}{\rho^2}\Big) \sum_{a=1}^N \omega_a Q^{(a)} .
\end{gather}
Imposing the rotational symmetry to the twisting vortex solutions
the conserved angular momentum of the vortex will be also conserved. 

Since the gauge field of our ansatz  has has no $x_3$ axis and 
the $x_3$ dependent phases in  scalar field are compensated 
by a gauge transformation, the solution has a translation invariance 
along $x_3$ axis. The associated  conserved momentum (per 
unit length) is given by
\begin{align}
P = \int d^2x T^{0}_z.
\end{align}

Our twisting solution has the nontrivial longitudinal current density 
flowing through  any transverse plane of the vortex. Integrating 
the Eq.~(\ref{j0eq}) over the $(x_1,x_2)$ plane shows that total 
current, and total charge density per unit vortex length are both 
zero,
\begin{align}
I_3 = \int d^2x J_3 =- \int d^2x J_0 = - I_0 =  0.
\end{align}

\section{Summary}

In this paper we have presented twisted BPS vortex solutions 
in the extended Abelian Higgs model where the scalar field 
is in the fundamental representation of {SU(N+1)} group.  
This twisted vortex solution has different phases in each 
components of scalar multiplet. The phases are linear in 
$x_3$ and $t$. Twisting static vortex introduces an additional 
energy contributions which are related with the conserved 
Noether charge of the remaining symmetry. They have 
the conserved longitudinal momentum which can be 
expressed by the Noether charge. 

It will be interesting to check whether it is possible to support 
the traveling wave of the form $e^{i (a t \mp b x_3)}$ with 
$a^2 - b^2 \ne 0$ in the current model.

From the mathematical point of view our vortex solutions 
are interesting. Moreover, from the physical point of view 
they have potentially important applications in various areas 
of physics, in high energy physics, cosmology, and condensed 
matter physics. In particular they could have important roles 
in multi-gap superconductors and multi-component 
Bose-Einstein condensates. This is because they are a natural 
generalization of the Abrikosov vortex in Ginzburg-Landau 
model of superconductor.  

{\bf ACKNOWLEDGEMENT}

~~~The work is supported in part by the National Research 
Foundation of Korea funded by the Ministry of Education 
(Grants 2015-R1D1A1A0-1057578 and 2015-R1D1A1A0-1059407), 
and by Konkuk University.

\end{document}